\documentclass{article}
\usepackage{spconf,amsmath,graphicx, amssymb, mathrsfs, subfig, multirow}

\usepackage{enumitem}
\setlist{nosep, leftmargin=14pt}
\usepackage{url}
\usepackage{mwe} % to get dummy images
\usepackage{threeparttable} % footnote for tables

\title{Neural Radiance Projection}
\name{
\parbox{\linewidth}{\centering
Pham Ngoc Huy$^{\dagger}$, 
and Tran Minh Quan$^{\dagger\ddagger}$ \\
Email : \{hpham, qtran\}@talosix.com 
}%
}%

\address{
\parbox{\linewidth}{\centering
$^{\dagger}$Talosix, Ho Chi Minh City, Vietnam \qquad
$^{\ddagger}$VinUniversity, Ha Noi, Vietnam 
}
}

\begin{document}
%\ninept
%
\maketitle
\begin{abstract}
The proposed method, Neural Radiance Projection (NeRP), addresses the three most fundamental shortages of training such a convolutional neural network on X-ray image segmentation: dealing with missing/limited human-annotated datasets; ambiguity on the per-pixel label; and the imbalance across positive- and negative- classes distribution. 
By harnessing a generative adversarial network, we can synthesize a massive amount of physics-based X-ray images, so-called Variationally Reconstructed Radiographs (VRRs), alongside their segmentation from more accurate labeled 3D Computed Tomography data.
As a result, VRRs present more faithfully than other projection methods in terms of photo-realistic metrics. 
Adding outputs from NeRP also surpasses the vanilla UNet models trained on the same pairs of X-ray images. 
\end{abstract}
\section{Introduction}
\label{sec:intro}
X-ray (XR) imaging can create photos of the body inside. 
The images show the parts in different intensities of grayscale. 
It is because various tissues absorb particular amounts of radiation. 
The most typical use of X-rays is checking for broken bones (fracture), but X-rays are also used for other purposes: 
chest XRs can spot pneumonia, or mammograms use XRs to look for breast cancer. 
The amount of radiation from an XR is small. 
For example, a chest XR gives out a radiation dose similar to the amount our body naturally exposes to from the environment over ten days.

Although XR imaging is widely used as an indispensable tool in diagnostic medicine thanks to its fast and cheap operations, the collected XR datasets are often long-tail and imbalanced, i.e., the number of negative cases is dominant compared to those that have disease-detection, because many typical cases take regular healthcare checkup annually. 
There is also a worldwide shortage of fully understanding the image interpretation since many overlapping things are projected on the same pixels. 
In order to resolve this issue, more advanced imaging techniques such as Computed Tomography (CT) or Magnetic Resonance (MR) are solicited to disentangle the ambiguity in 3D space. 
They provide more reliable and golden ground truth to examine the patients. 
However, only severely positive cases need to take further scans due to their expense in cost and time operations. 
On the other hand, Artificial Intelligence and Machine Learning have shown remarkable performance recently in the automated evaluation of medical images. 
Nevertheless, they require well-annotated, balanced, and relatively large datasets to produce an excellent supervised-learning model for deployment. 
These drawbacks pose significant challenges in starting an XR-based medical image analysis study for new pathology problems when the requirements from both sides burden the others. 

Our work stems from a recently advanced solution on Neural Radiance Field (NeRF)~\cite{mildenhall2020nerf} that attempts to synthesize/reconstruct the scene from sparse-view images. 
We leverage the ray marching and ray sampling strategy to implement our differentiable Neural Radiance Projection (NeRP) to deal with common issues which hinder the deep learning approach on direct and end-to-end solutions for radiograph image analysis:
\textbf{(1)} Missing/Limited amount of data: Usually, when we start a brand new image analysis problem with a specific goal, we are entirely lacking the data or having only a few samples for particular tasks. 
\textbf{(2)} Annotations often come with uncertainties: Due to the blurry effect and depth estimation, it is ambiguous to label the per-pixel region of interest, especially on projected radiographs when many things can accumulate on the same pixels. 
\textbf{(3)} Imbalanced data since positively severe cases rarely appear in the datasets: in our case, less positive images for certain classes in XR collection can be readjusted by more positive CT scans. 
% \begin{itemize}
%  \item Missing/Limited amount of data.
%  \item Annotations often come with uncertainty due to the blurry effect and depth estimation. 
%  \item Imbalanced data since positively severe cases rarely appear in the datasets. 
% \end{itemize}
%

In summary, the proposed method aims to produce many more physics-based XR images with golden CT scan labels. 
Our contributions are several-fold as follows:
\begin{itemize}
  \item Variational Cameras within proximities to generate a large number of labeled pairs of XR images. 

  \item A Generative-Adversarial Network (GAN) to enhance the realism of generated XR data. 

  \item A data augmentation to enrich the limited/missing training pair of XR images. 
\end{itemize}
To the best of our knowledge, this work is the first attempt that leverages heterogeneous data from both CT and XR to train a segmentation model and improve its performance compared to the baselines of using XR data only. 
We demonstrate its usability in real-world applications of segmenting X-ray images.

\section{Method}
\label{sec:method}
\subsection{XR-like image generation}

\begin{figure}[tp]
% \centering
\subfloat[]{
\centering
\includegraphics[width=0.38\linewidth]{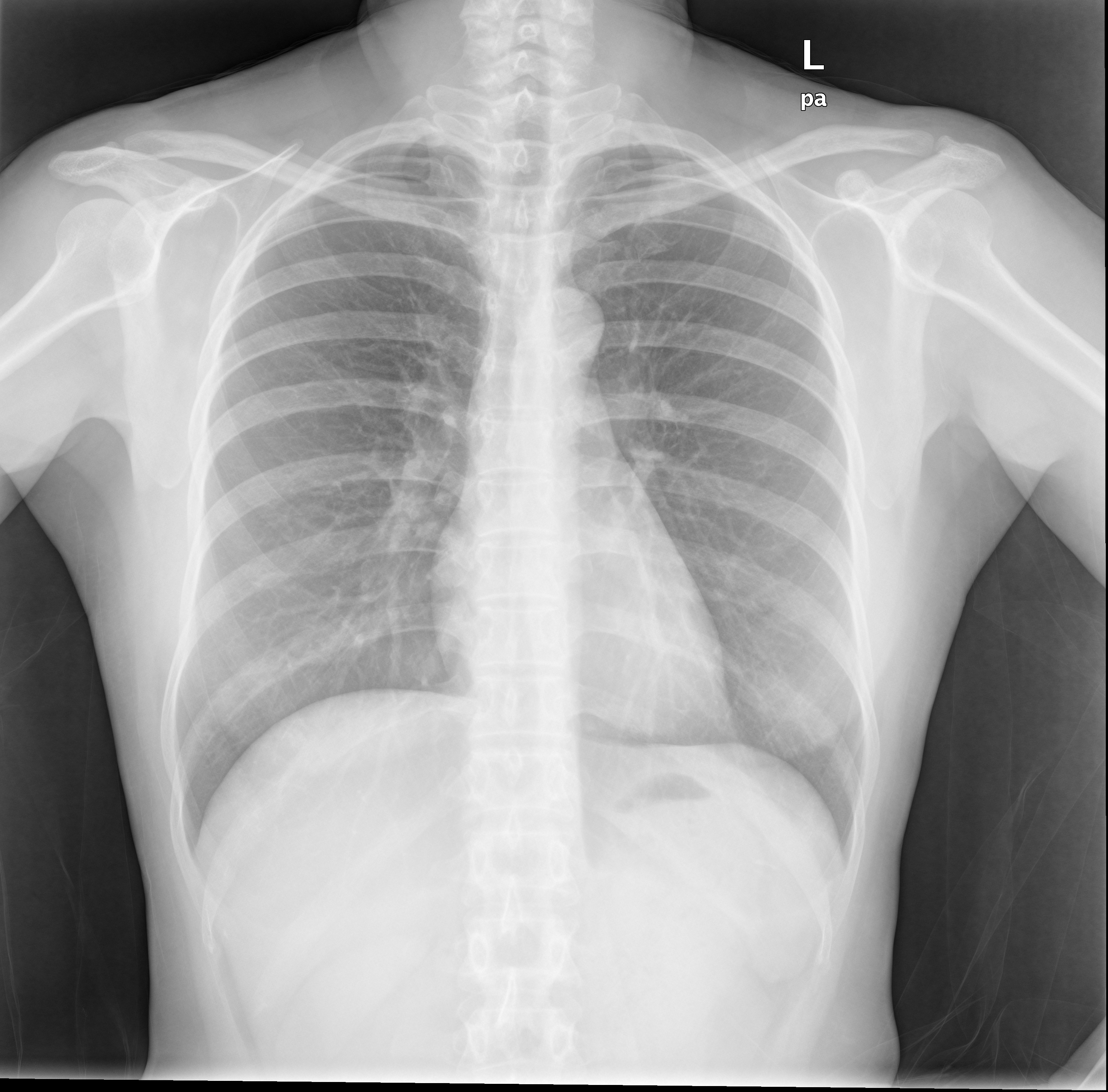}
}
\subfloat[]{
\centering
\includegraphics[width=0.62\linewidth]{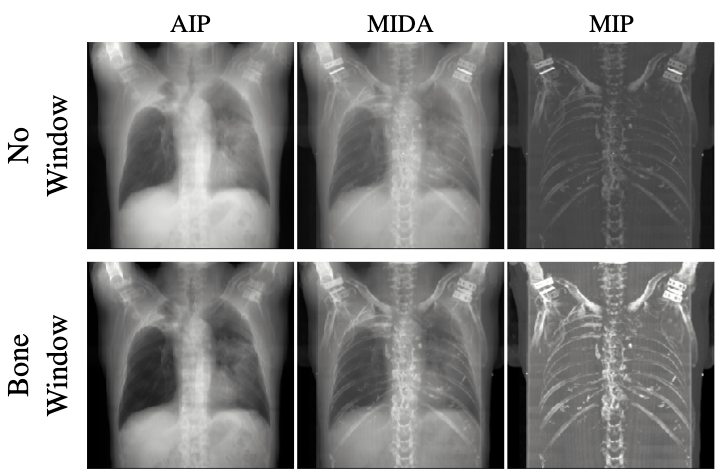}
}
\caption{(a) A real XR image (b) XR-like images generated by DDR methods with or without pre-windowing.}
\label{fig:drr}
\end{figure}

The most direct way to transform a typical 3D CT volume data to a 2D image is to average the intensity (ray accumulation) along particular directions (view direction) to generate such an X-ray-like image.  
This approach gives us a brief sense of how overall the 3D CT data looks like in the context of projection, but it does not look similar to the real X-ray image that is commonly taken, in terms of contrast or intensity, even with common pre-windowing setups of CT data (see Fig.~\ref{fig:drr}). 
%but its fine-detail of image features appear to be blurred.
%
It can be explained that this averaging intensity projection (AIP) does not fully reflect the photon attenuation of heterogeneous absorber (CT data) measuring in the Hounsfield Unit (HU) scale. 
Several techniques such as MIDA~\cite{hussain2017differential} combines a linear lookup table (intensity versus opacity) and non-linear ones such as Maximum-Intensity Projection (MIP) to promote better certain material visualization. 

The 2D Digitally Reconstructed Radiographs (DRRs) generated from 3D CT images by more advanced methods such as volume rendering algorithms~\cite{Siddon1985, jacobs1998fast}, have better representations of photo-realistic projections. 
There is also seminal work that attempts to generate physics-based XR such as DeepDRR~\cite{deepdrr}, XraySyn~\cite{xraysyn}.
They both require hand-crafted designed matters of interest (masks) to determine material thickness criteria to discriminate different radiant interactions. 
On the other hand, XPGAN~\cite{xpgan} produces non-physics-based XR with source location-free.  
However, even if it can approximate the projection by a 3D-2D neural network, the novel view synthesis is not amendable since the viewpoint does not involve in the parameters of training. 
In an extreme case, slightly rotating the CT volume results in a very different XR image. 
Therefore, it motivates us to develop a fully end-to-end learning method to create a photo that looks realistic to the particular XR images without human assistance of segmentation or prior knowledge of radiance interaction but still be able to manipulate the viewpoint. 

\subsection{Model overview}

\begin{figure*}[ht]
\centering
\includegraphics[width=\textwidth]{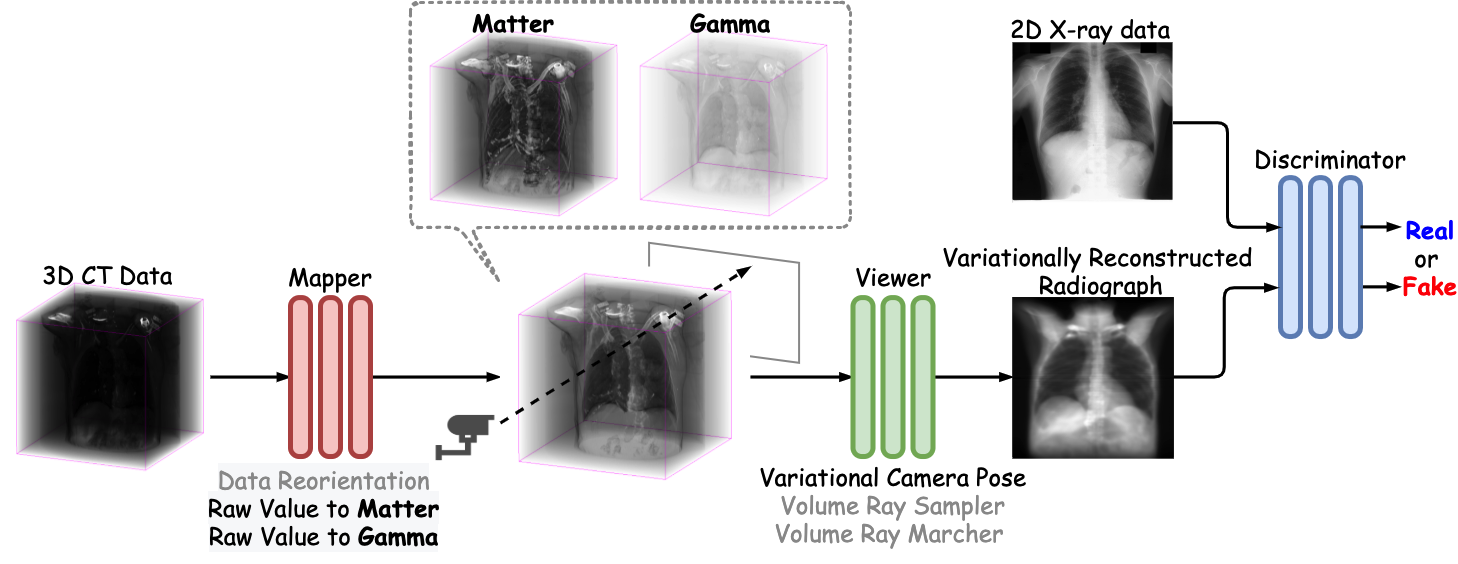}
\caption{Neural Radiance Projection Overview.}
\label{fig:nerp}
\vspace{-4mm}
\end{figure*}

%
% Therefore, it motivates us to develop a learning method to produce a two-channel volume that consists of a Matter Field ($\mu$ volume) combined with an Opacity Field ($\gamma$ volume) from the original 3D CT data, in which after the simulated x-ray beams penetrate through, it will produce a photo that looks realistic to the particular XR images. 
%
As shown in Fig.~\ref{fig:nerp}, the proposed method, NeRP, consists of these building blocks: a Generator $\mathcal{G}$ includes a Mapper $\mathcal{M}$, a Viewer $\mathcal{V}$; and a Discriminator $\mathcal{D}$ that are parameterized by following objective functions:

\begin{equation}
    \min_{ \theta_{\mathcal{G}}}\max_{\theta_{D}} \quad L_{adv}(\mathcal{G}, \mathcal{D}) + \lambda L_{reg}(\mathcal{G})
    \label{eq:major}
\end{equation}

\begin{equation}
    \min_{ \theta_{\mathcal{M}}, \theta_{\mathcal{V}} }\max_{\theta_{D}} \quad L_{adv}(\mathcal{M}, \mathcal{V}, \mathcal{D}) + \lambda L_{reg}(\mathcal{M}, \mathcal{V})
    \label{eq:minor}
\end{equation}

The Mapper $\mathcal{M}$ takes a single channel 3D CT volume $y_{1 \times D \times H \times W} \sim P_{Y}$, which has been normalized to produce a two-channel volume $y^{\mu\gamma}_{2 \times D \times H \times W}$ that consists of a Matter Field ($y^{\mu}_{1 \times D \times H \times W}$ volume) combined with an Opacity Field ($y^{\gamma}_{1 \times D \times H \times W}$ volume). 
The subscripts indicate the shapes of tensors which have channel $C=\{1,2\}$, depth $D$, height $H$, and width $W$, respectively.
They will be omitted for simplicity. 
The maximum bandwidth of $y^{\gamma}$ is set \textbf{100-fold smaller} than $y^{\mu}$ to foster the XR-like effect and discriminate different roles between them. 
Otherwise, these two channels are interchangeable and behave similarly to each other.
Then the Viewer $\mathcal{V}$ accepts the radiance $y^{\mu\gamma}$, constructs the beam emission absorption model with ray bundle $r\sim{}P_{R}$ so that these rays can interact with the implicit volume function $y^{\mu}$ and the opacity $y^{\gamma}$. 
The proposed method also randomly samples a (variationally) perspective camera pose $c\sim{}P_{C}$ within a certain proximity of Field of View (FoV) to form the final image on the screen. 
The result is therefore called Variationally Reconstructed Radiographs (VRRs), as distinguish to DRRs. 
We employ Generative Adversarial Networks (GANs)~\cite{goodfellow2014generative} in XPGAN~\cite{xpgan} to treat the VRR results as fake images and train the networks adversarially with the true XR samples that drawn from the real datasets $x\sim{}P_{X}$. 
The adversarial loss $L_{adv}$ is formally defined in Eq.~\ref{eq:adv} with the expectation of NeRP result is expanded in Eq.~\ref{eq:exp}:

\begin{equation}
\begin{aligned}
   L_{adv}(\mathcal{G}, \mathcal{D}) = \mathop{\mathbb{E}}_{x \sim P_{X}}\left[\log \mathcal{D}(x)\right] + \mathop{\mathbb{E}}_{y \sim P_{Y} }\left[1-\log \mathcal{D}(\mathcal{G}(y))\right]
    % = \mathop{\mathbb{E}}_{x \sim P_{X}}\left[\log D(x)\right] + \mathop{\mathbb{E}}_{\substack{y \sim P_{Y} \\ r \sim P_{R} \\ c \sim P_{C}}}\left[1-\log D(G(y)), r, c\right] 
\end{aligned}
\label{eq:adv}
\end{equation}

\begin{equation}
\begin{aligned}
   \mathop{\mathbb{E}}_{y \sim P_{Y} }\left[1-\log \mathcal{D}(\mathcal{G}(y))\right]
    =  \mathop{\mathbb{E}}_{\substack{y \sim P_{Y} \\ r \sim P_{R} \\ c \sim P_{C}}}\left[1-\log \mathcal{D}(\mathcal{V}(\mathcal{M}(y), r, c))\right]
\end{aligned}
\label{eq:exp}
\end{equation}

 %We add a per-pixel matching loss between the generated image and the ray tracing rendered image to constraint the projected images to have the same anatomical structure as of its original CT volumes.  
Last but not least, we add an $\ell_1$ per-pixel loss between the generated image and the ray tracing rendered version, weighted by $\lambda$ to regularize the projected images to have the same anatomical structure as of its original CT volumes.  

\begin{equation}
    L_{reg}(\mathcal{G}) = \mathop{\mathbb{E}}_{y \sim P_{Y}}\left[\lVert \mathcal{G}(y) - \mathcal{P}roj(y) \rVert_1\right]
\end{equation}
For wide-range choices of $\{0.05, 0.1, 0.5, 1, 5, 10\}$, $\lambda=1$ yields better convergence and image quality compared to other options. 

\subsection{Implementation details}
To leverage the power of state-of-art models, we utilize architectures of the current latest methods. 
Specifically, we use V-Net~\cite{milletari2016v} for Mapper $\mathcal{M}$ and U-Net~\cite{ronneberger2015u} for Discriminator $\mathcal{D}$~\cite{schonfeld2020u}. 
It is worth noting that having U-Net as a Discriminator stabilizes the training process. 
For the Viewer $\mathcal{V}$, the ray bundle is set as the same as image size ($256 \times 256$), each ray is sampled with 512 points. 
The entire system is trained for 200 epochs with an NVIDIA 3080 Ti GPU for one week. 
 
%

%%%%%%%%%%%%%%%%%%%%%%%%%%%%%%%%%%%
% CT Histogram under many windows %
%%%%%%%%%%%%%%%%%%%%%%%%%%%%%%%%%%%
% \begin{figure}[h]
%     \centering
%     \includegraphics[width=0.45\textwidth]{fig/ct_slice_windowing.png}
%     \caption{CT Histogram under many windows}
%     \label{fig:ct_hist}
% \end{figure}
%%%%%%%%%%%%%%%%%%%%%%%%%%%%%%%%%%%
% End                             %
%%%%%%%%%%%%%%%%%%%%%%%%%%%%%%%%%%%

%%%%%%%%%%%%%%%%%%%%%%%%%%%%%%%%%%%
% CT Histogram under many windows %
%%%%%%%%%%%%%%%%%%%%%%%%%%%%%%%%%%%
% \begin{figure}[h]
%     \centering
%     \includegraphics[width=0.45\textwidth]{fig/projection_vs_windowing.png}
%     \caption{Projection under many windows}
%     \label{fig:projection_windows}
% \end{figure}
%%%%%%%%%%%%%%%%%%%%%%%%%%%%%%%%%%%
% End                             %
%%%%%%%%%%%%%%%%%%%%%%%%%%%%%%%%%%%

\section{Data}
\label{sec:data}
\begin{table}[h]
\centering
\vspace{-5mm}
\caption{Statistics of CT volumes and XR images.} %\\ used in the experiments}
\begin{tabular}{|c|c|c|r|}
% \vspace{-10mm}
\hline
\textbf{Type}        & \textbf{Dataset}               & \textbf{Description}  & \textbf{Total} \\ \hline
\multirow{3}{*}{CT}  & NSCLC \cite{nsclcDataset}      & Lung Cancer  & 402 \\ \cline{2-4} 
                     & MosMed \cite{morozov2020mosmeddata} & COVID-19 & 1110          \\ \cline{2-4} 
                     & \multicolumn{2}{c|}{Total}                            & \textbf{1512} \\ \hline
\multirow{4}{*}{CXR} & ChinaSet \cite{twoChestXray}   & Tuberculosis & 566          \\ \cline{2-4} 
                     & Montgomery \cite{twoChestXray} & Tuberculosis & 138          \\ \cline{2-4} 
                     & JSRT \cite{jsrtDataset}        & Lung nodules & 247          \\ \cline{2-4} 
                     & \multicolumn{2}{c|}{Total}                            & \textbf{951} \\ \hline
\end{tabular}
\label{tab:datasetStat}
\end{table}

The CT volumes were prepared from two datasets. The first is Non-Small Cell Lung Cancer CT Volume, comes from the radio-genomic database of non-small cell lung cancer patients at Stanford University Medical Center and Palo Alto Veterans Affairs Healthcare System. This collection has 402 annotated CT from a cohort of 211 patients. The MosMed dataset is comprised of COVID-19 related CT scans from the hospitals in Moscow, Russia.
Besides that, the chest X-Ray images come from three different sources. 
Both Montgomery and ChinaSet were publicized by the US National Library of Medicine in 2014. The former dataset comes from the Tuberculosis screening program of Montgomery County, Maryland, USA, and consists of 138 anterior-posterior CXR images; the latter contains 566 frontal projections in both normal and tuberculosis patients~\cite{twoChestXray}. %
The Japanese Society of Radiological Technology published the JSRT database in 1998. This source includes 247 chest radiographs from Japanese and American institutions. These images were used to identify the pulmonary nodules in terms of five degrees~\cite{jsrtDataset}. 
The dataset statistics are collected in the \text{Table \ref{tab:datasetStat}}.

\section{Results}
\label{sec:result}
\subsection{Quality of generated images}

\begin{table}[htbp]
\vspace{-5mm}
\caption{Image quality metrics:\\
Higher IS and lower FID, KID mean better image quality.}
\label{tab:inception}
\centering
\begin{tabular}{|c|c|c|c|}
\hline
\textbf{Method} & \textbf{IS}~$\uparrow$ & \textbf{FID}~$\downarrow$ & \textbf{KID}~$\downarrow$ \\ \hline
DRR~\cite{Siddon1985}& 1.799 $\pm$ 0.050 & 197.071 & 0.226 $\pm$ 0.013 \\ \hline
DeepDRR~\cite{deepdrr} & 2.323 $\pm$ 0.062 & \textbf{153.399} & \textbf{0.158 $\pm$ 0.013} \\ \hline
NeRP    & \textbf{2.454 $\pm$ 0.072} & 161.675 & 0.161 $\pm$ 0.013 \\ \hline
\end{tabular}
\end{table}

% The Frechet Inception Distance (FID)~\cite{??} score, 
Many relevant metrics can can be used to evaluate the results on how photo-realistic they are over the real datasets: such as Inception Score (IS)~\cite{salimans2016improved}, Frechet Inception Distance (FID) score~\cite{heusel2017gans} and Kernel Inception Distance (KID) score~\cite{binkowski2018demystifying}. 
We pre-generate 4020 XR images from 402 CT data from NLCSC (one volume produces 10 projections with different viewpoints) for each setup (DRR, DeepDRR, NeRP) and compare against each other over the distribution of Chest XR datasets~\cite{twoChestXray, jsrtDataset}. 
Table~\ref{tab:inception} shows the results of each metrics. 
Our method has similar scores to DeepDRR, and both are better than DRR. 
Note that although DeepDRR has more parameters to control, such as spectrum, photon counts, or scattering, it does not support generating the projected labels. 

%%%%%%%%%%%%%%%%%%%%%%%%%%%%%%
% FID, KID, IS metrics table %
%%%%%%%%%%%%%%%%%%%%%%%%%%%%%%
% This table compares metrics of NeRP and AIP/DRR
% Code for calculation 
% https://colab.research.google.com/drive/1KI1zb-fAMXR08TDeRilSsqncd1cAUz-i?usp=sharing
% \begin{table}[h]
% \centering
% \caption{Image quality metrics \tnote{1}}
% \vspace{-3mm}
% \begin{threeparttable}
% \begin{tabular}{|l|c|c|r|c|r|}
% \hline
% \multicolumn{1}{|c|}{\multirow{2}{*}{\textbf{Method}}} & \multirow{2}{*}{\textbf{FID~$\downarrow$}}  & \multicolumn{2}{c|}{\textbf{KID}} & \multicolumn{2}{c|}{\textbf{IS}} \\ \cline{3-6} 
% \multicolumn{1}{|c|}{} & & Mean & \multicolumn{1}{c|}{SD} & Mean & \multicolumn{1}{c|}{SD} \\ \hline
% DRR & \multicolumn{1}{r|}{207.8176} & \multicolumn{1}{r|}{0.2465} & 0.0180 & \multicolumn{1}{r|}{1.7994} & 0.0737 \\ \hline
% NeRP & \multicolumn{1}{r|}{\textbf{170.6253}} & \multicolumn{1}{r|}{\textbf{0.1833}} & 0.0177 & \multicolumn{1}{r|}{\textbf{2.4687}} & 0.0491 \\ \hline
% \end{tabular}
% \begin{tablenotes}
% \item Lower FID, KID and higher IS mean better image quality.
% \end{tablenotes}
% \end{threeparttable}
% \end{table}

%%%%%%%%%%%%%%%%%%%%%%%%%%%%%%
% End                        %
%%%%%%%%%%%%%%%%%%%%%%%%%%%%%%

\subsection{Lung Segmentation}
\subsubsection{Missing annotated XR data}
\begin{figure}[ht]
    \centering
    \includegraphics[width=0.5\textwidth]{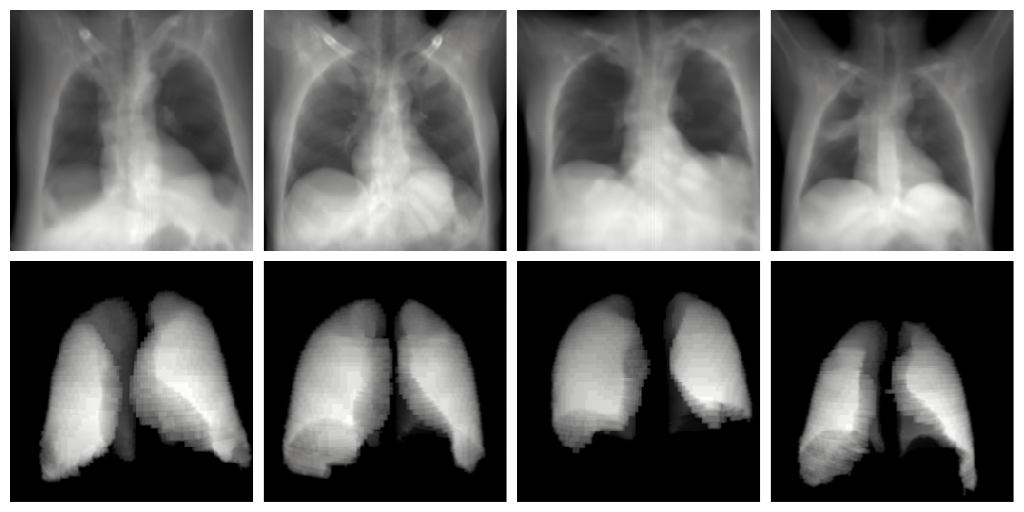}
    \caption{Samples of NeRP pairs for Lung Segmentation task.}
    \label{fig:sample_nerp}
    \vspace{-5mm}
\end{figure}

In order to show the advantages of NeRP, we compare its performance against DRRs on Lung segmentation of Chest XR in the scenario of lacking the training data entirely, only test on given manual sets as discussed. 
To make a fair assessment, we set the same viewpoint-generated VRRs from NeRP and DRRs. 
The only difference between them is that our radiance $y^{\mu\gamma}$ is learnable while DRRs have fixed HU-based $y^{\mu}$ and constant $y^{\gamma}$. 
Fig.~\ref{fig:sample_nerp} indicates some VRR samples generated by NeRP that have been used for training a vanilla UNet to perform pixel-based lung segmentation problems. 
As can be seen in Table~\ref{tab:drrnerp}, using our method can outperform a DRR-based dataset up to 5\% on the Dice Score~\cite{zou2004statistical} metrics. 
This can be explained that the XR images produced by NeRP are much typical than by DRRs, and hence require less overhead transfer compared to those direct DRR CT-generated samples. 

%%%%%%%%%%%%%%%%%%%%%%%%%%%%%%
% Sample of generated images %
%%%%%%%%%%%%%%%%%%%%%%%%%%%%%%
%%%%%%%%%%%%%%%%%%%%%%%%%%%%%%
% End                        %
%%%%%%%%%%%%%%%%%%%%%%%%%%%%%%

%%%%%%%%%%%%%%%%%%%%%%%%%%%%%%%%%%%%%%%%%
% Dice score table for Lung segmentation%
% using only generated image            %
%%%%%%%%%%%%%%%%%%%%%%%%%%%%%%%%%%%%%%%%%
% \begin{table}[htbp]
% \centering
% \caption{Comparison of training DRR- and NeRP-generated pairs from 3D CT data, and test on human-annotated 2D XR images for Lung Segmentation}
% \begin{tabular}{|c|c|c|}
% \hline
% \multirow{2}{*}{\textbf{Dataset}} & \multicolumn{2}{c|}{\textbf{Dice Score}} \\ \cline{2-3} 
%                          & DRR            & NeRP          \\ \hline
% CHINASET                 & 0.8033         & \textbf{0.8491}        \\ \hline
% MONTGOMERY               & 0.8229         & \textbf{0.8357}        \\ \hline
% JSRT                     & 0.8060         & \textbf{0.8261}        \\ \hline
% Average                  & 0.8068         & \textbf{0.8412}        \\ \hline
% \end{tabular}
% \label{tab:drrnerp}
% \end{table}

\begin{table}[htbp]
\centering
\caption{Comparison of Dice Scores of DRR- and NeRP-generated pairs from 3D CT data, and test on human-annotated 2D XR images for Lung Segmentation.}
\vspace{-3mm}
\begin{tabular}{|c|c|c|c|c|}
\hline
% \multirow{2}{*}{\textbf{Dataset}} & CHINA & MONTGO & \multirow{2}{*}{JSRT} & \multirow{2}{*}{Average} \\
%  & SET & MERY &  &  \\ \hline
\textbf{Dataset} & ChinaSet & Montogomery & JSRT & Average \\ \hline
DRR~\cite{Siddon1985} & 0.8033 & 0.8060 & 0.8229 & 0.8068 \\ \hline
NeRP & \textbf{0.8491} & \textbf{0.8261} & \textbf{0.8357} & \textbf{0.8412} \\ \hline
\end{tabular}
\label{tab:drrnerp}
\vspace{-5mm}
\end{table}

%%%%%%%%%%%%%%%%%%%%%%%%%%%%%%%%%%%%%%%%%
% End                                   %
%%%%%%%%%%%%%%%%%%%%%%%%%%%%%%%%%%%%%%%%%

\subsubsection{Limited annotated XR data}

%%%%%%%%%%%%%%%%%%%%%%%%%%%%%%
% Plot of Prediction         %
%%%%%%%%%%%%%%%%%%%%%%%%%%%%%%
\begin{figure}[htbp]
    \vspace{-1mm}
    \centering
    \includegraphics[width=0.5\textwidth]{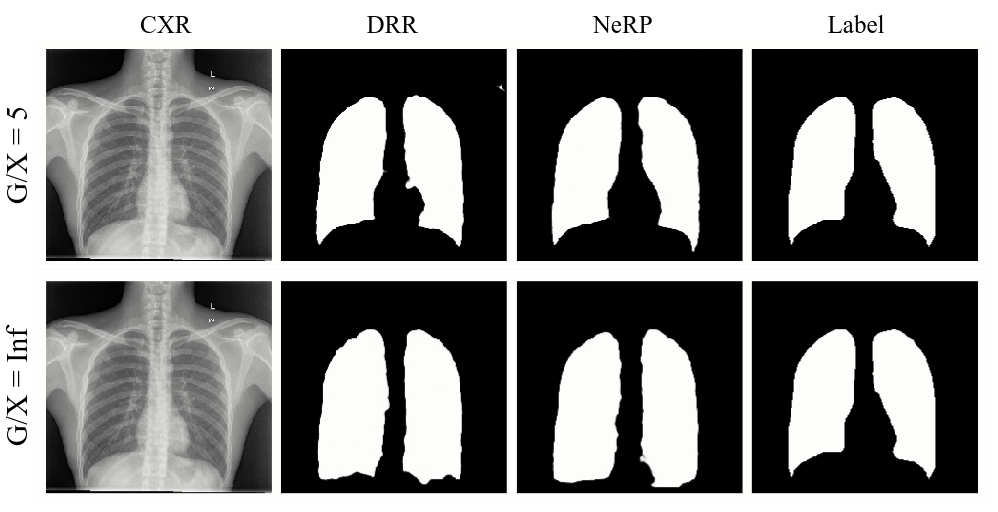}
    \vspace{-8mm}
    \caption{Samples of NeRP data for Lung Segmentation task.}
    \vspace{-5mm}
    \label{fig:samplelung}
\end{figure}
%%%%%%%%%%%%%%%%%%%%%%%%%%%%%%
% End                        %
%%%%%%%%%%%%%%%%%%%%%%%%%%%%%%

%%%%%%%%%%%%%%%%%%%%%%%%%%%%%%
% Plot of Dice score         %
%%%%%%%%%%%%%%%%%%%%%%%%%%%%%%
\begin{figure}[ht]
    \vspace{-5mm}
    \centering
    \includegraphics[width=\linewidth]{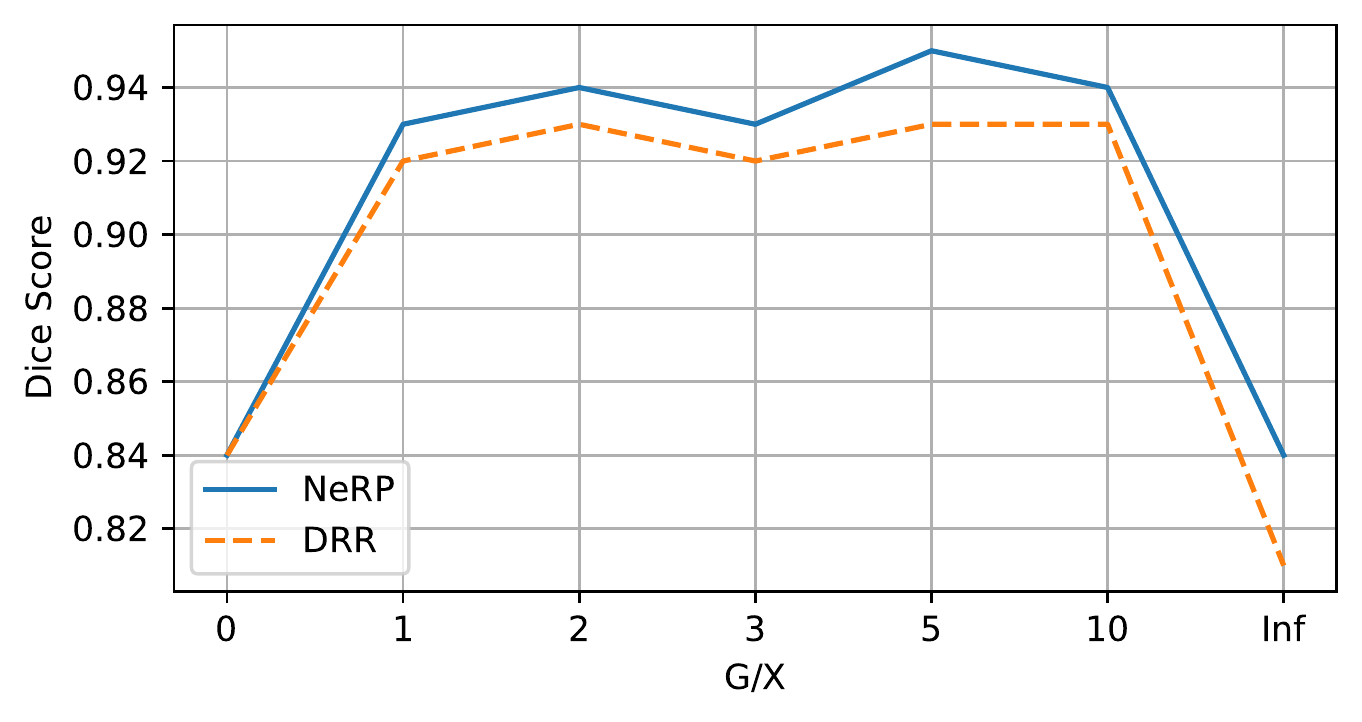}
    \vspace{-9mm}
    \caption{Dice scores of Lung Segmentation task.}
    \vspace{-3mm}
    \label{fig:dice_score_lung}
\end{figure}
%%%%%%%%%%%%%%%%%%%%%%%%%%%%%%
% End                        %
%%%%%%%%%%%%%%%%%%%%%%%%%%%%%%

%%%%%%%%%%%%%%%%%%%%%%%%%%%%%%%%%%%%%%%%%
% Pleural Effusion Lung Collapse         %
%%%%%%%%%%%%%%%%%%%%%%%%%%%%%%%%%%%%%%%%%
\begin{figure}[!]
    \centering
    \includegraphics[width=0.5\textwidth]{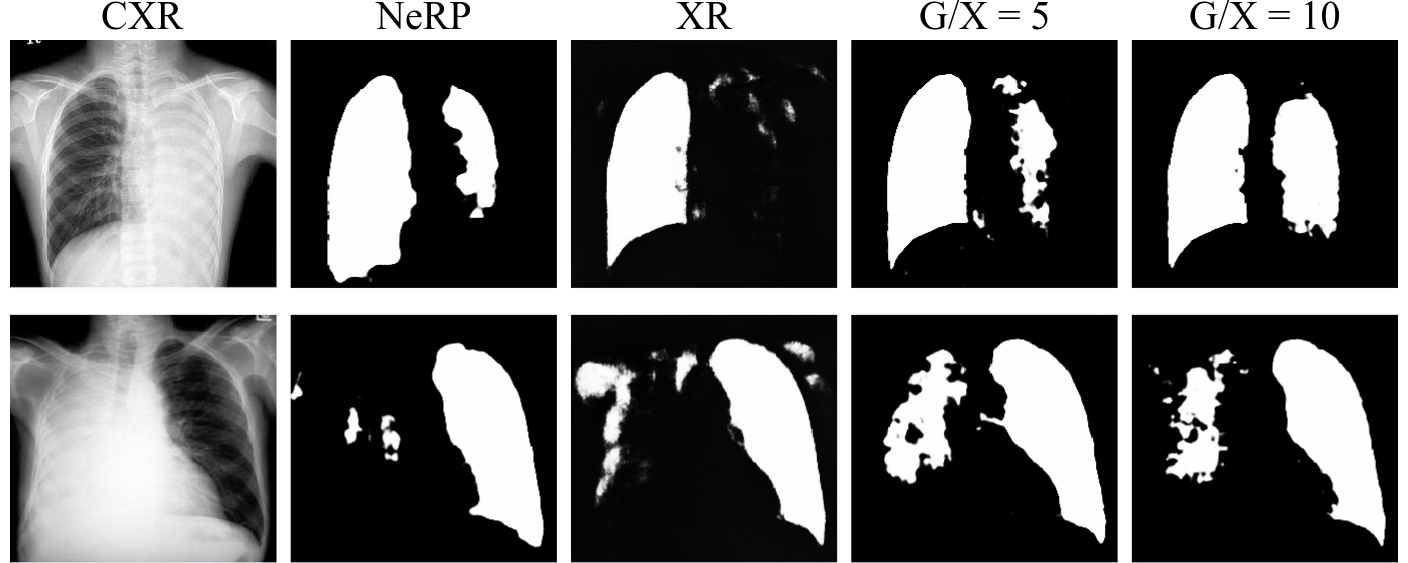}
    \vspace{-5mm}
    \caption{Comparison in Pleural effusion lung collapse samples.}
    \label{fig:pleural_effusion}
    \vspace{-5mm}
\end{figure}
%%%%%%%%%%%%%%%%%%%%%%%%%%%%%%
% End                        %
%%%%%%%%%%%%%%%%%%%%%%%%%%%%%%

\noindent
In this assessment, we gradually include portions of Chest XR with Lung annotation to train the U-Net and observe the performance over the quota of having extra annotated subsets of 2D XR data (see Fig.~\ref{fig:samplelung}). 
Fig.~\ref{fig:dice_score_lung} shows the Dice Scores on the separated test set for this setup. %
The horizontal axis represents the ratio (\textbf{G/X}) between the number of the generated images and XR images: $0$ means that the model was trained from XR only, and \textit{Inf} means that model was trained from the generated images only. 
The results indicate that mixing the handful of annotated data with NeRP images yields better performance (up to 2\%) over conventional DRR.

Furthermore, we choose extreme cases of Chest XR images with Pleural Effusion Lung Collapse, i.e., lung lobe on one side is filled by high contrast and brightness pixel values. 
As depicted in Fig.~\ref{fig:pleural_effusion}, combining NeRP and XR data promotes better lung regions even if the appearance of dark areas is partially missing. 

%%%%%%%%%%%%%%%%%%%%%%%%%%%%%%%%%%%%%%%%%
% Dice score table for Lung segmentation%
% with many ratios.                     %
%%%%%%%%%%%%%%%%%%%%%%%%%%%%%%%%%%%%%%%%%
% \begin{table}[htbp]
% \centering
% \caption{Dice Score of Lung Segmentation}
% \vspace{-3mm}
% \begin{threeparttable}
% \begin{tabular}{|l|c|c|c|c|c|c|c|}
% \hline
% \textbf{G/X}\tnote{1}  & 0    & 1    & 2    & 3    & 5    & 10   & Inf  \\ \hline
% \textbf{DRR}  & 0.84 & 0.92 & 0.93 & 0.92 & 0.93 & 0.93 & 0.81 \\ \hline
% \textbf{NeRP} & 0.84 & \textbf{0.93} & \textbf{0.94} & \textbf{0.94} & \textbf{0.95} & \textbf{0.94} & \textbf{0.84} \\ \hline
% \end{tabular}
% \begin{tablenotes}
% \item[1] Ratio between the number of the generated images and XRay images. 0 means that model was trained from XRay only and Inf means that model was trained from the generated images only.
% \end{tablenotes}
% \end{threeparttable}
% \end{table}
%%%%%%%%%%%%%%%%%%%%%%%%%%%%%%%%%%%%%%%%%
% End                                   %
%%%%%%%%%%%%%%%%%%%%%%%%%%%%%%%%%%%%%%%%%

\section{Conclusion}
\label{sec:conclusion}
We present NeRP, which can produce VRRs. 
Comparing to conventional DRRs, NeRP can generate much more faithful images in common photo-realistic metrics. 
This work addresses the three most fundamental issues of establishing a machine learning model to solve XR-based image analysis problem: \textbf{(1)} Missing/Limited amount of data; \textbf{(2)} Uncertain annotation; and \textbf{(3)} Imbalanced datasets.   
By leveraging the ability of massive amount pairs of segmentation data created by NeRP, we can utilize more golden ground truth and ready-to-use CT data to balance the currently long-tail datasets with more accurate projected 3D labels. 
This usefulness will be applied for more XR-based image analysis problems such as rare cancer detection, diagnosis, and prognosis.

% \vspace{-2mm}
% %
% We present XPGAN in training a deep neural network to classify CXR images, targets to COVID-19 detection, with limited labeled data. 
% %
% The results show that the F1 score from the XPGAN improves up to $\sim$2\% over the baselines, which has the same architecture of classifier (DenseNet121). 
% %
% % The proposed method helps to detect COVID-19 patients in early-stage and prioritize isolating decision making. 
% %
% Thanks to the nature of the generative model, we can synthesize the CXR images from confirmed cases of CT data. 
% %
% In future work, we plan to have an in-depth study of jointly training both 2D and 3D model to improve XPGAN. 

\section{Compliance with Ethical Standards}
\label{sec:ethics}
% This is a retrospective study of the COVID-19 pandemic using data in Vietnam and the public domain. 
% %
% The study was performed in line with the principles of the Declaration of Helsinki. 
% %
% Approval was granted by the Ministry of Health, Vietnam (Date: April 15 2020, No. 1724/QD-BYT).

All procedures performed in studies involving human participants were in accordance with the ethical standards of the institutional and/or national research committee and with the 1964 Helsinki declaration and its later amendments or comparable ethical standards.
For this type of retrospective study, formal consent is not required. 
% This research study was conducted retrospectively using human subject data made available in open access. Ethical approval was *not* required as confirmed by the license attached with the open access data."

\section{Conflicts of Interest}
% \small
\label{sec:conflict}
The authors have no conflicts of interest to disclose. 
% This work was supported by VinBrain, a company funded by VinGroup, Vietnam. 
% Authors Tran Quy Tuong, Vu Minh Dien, Bui Van Giang and Bui Huu Trung have served on advisory boards for VinBrain. 
% All authors have no conflicts of interest to disclose. 
% %
% The authors would like to thank Nguyen Ngoc Hoang, Hoang Vu, Nguyen Nhat Linh, Nguyen Minh Man, Do Nam Phuong and the rest of VinBrain team, VinMec Hospital radiologists for kindly crowd-sourcing the data as well as providing the label annotations. 
%
% Thank Mr. Truc Thai for initiating ``why don't we coffee?" and led the team who is making a tremendous progress.   

\section{References}
% References should be produced using the bibtex program from suitable
% BiBTeX files (here: strings, refs, manuals). The IEEEbib.bst bibliography
% style file from IEEE produces unsorted bibliography list.
% -------------------------------------------------------------------------
\bibliographystyle{IEEEbib}
\bibliography{refs}

\end{document}